\documentclass[twocolumn,showpacs,preprintnumbers,amsmath,amssymb]{revtex4}
\usepackage{amssymb}
\usepackage[dvips]{graphicx}
\begin{document}

\title{Nernst effect in poor conductors and in the  cuprate superconductors}
\author{A. S. Alexandrov and V. N. Zavaritsky}

\affiliation{Department of Physics, Loughborough University, Loughborough LE11 3TU, United Kingdom}

\begin{abstract}
We calculate the Nernst signal in disordered conductors with the
chemical potential near the mobility edge. The Nernst effect
originates from interference of  itinerant and localised-carrier
contributions to the thermomagnetic transport. It reveals a strong
temperature and magnetic field dependence, which describes quantitatively the anomalous Nernst signal in high-$T_{c}$
cuprates. \end{abstract}


\maketitle

Thermomagnetic effects appear in conductors subjected to a
longitudinal temperature gradient ${\mathbf{\nabla}} T$ (in $x$
direction) and a perpendicular magnetic field $\mathbf{B}$ in $z$
direction. The transverse Nernst-Ettingshausen effect (further the
Nernst effect) is the appearance of a transverse electric field
$E_y$ in the third direction. This effect as well as the
longitudinal one were discovered by Nernst and Ettingshausen in a
bismuth plate in 1886 \cite{nernst}. The effect is known to be small in ordinary
metals. Indeed in the framework of a
single-band effective mass approximation it appears only in the
second order with respect to the degeneracy $k_BT/E_F \ll 1$ due
to a so-called Sondheimer cancellation \cite{sond}, if the
relaxation time $\tau(E)$ depends on energy. If $\tau$ does not
depend on energy, the Nernst signal disappears even for
nondegenerate carriers in the same approximation \cite {ANSE}.

Sufficiently large positive Nernst effect was found in high-$T_c$
cuprates in the vicinity of the resistive transition temperature
$T_c$ \cite{groups}. As in conventional superconductors it was
attributed to motion of vortices down the thermal gradient while a
small negative signal, measured well above $T_c$ \cite{clay}, was
ascribed to the relaxation time decreasing with carrier energy.
Such a negative signal may also originate from the counterflow of
carriers with opposite sign (the familiar ambipolar Nernst
effect), as  explained by a simple two band model for electrons
and holes with different mobilities \cite{lam}, and/or from a
charged density wave order \cite{rom}, as observed in
$NbSe_2$.

Recently much attention has been paid to the anomalously enhanced
\emph{positive} Nernst signal observed \emph{well above} $T_{c}$
in $La_{2-x}Sr_{x}CuO_{4}$ (LSCO-x) in a wide range of doping $x$
\cite{xu}. It has been attributed to the \emph{vortex} motion,
since the Sondheimer cancellation renders any 'normal state'
scenario allegedly implausible \cite{xu}. As a result, the magnetic
phase diagram of the cuprates has been revised with the upper
critical field $H_{c2}(T)$ curve not ending at $T_{c}$ but at a
much higher temperature \cite{wang1,wang2}. Most surprisingly,
Refs.\cite{wang1,wang2} estimated $H_{c2}$ \emph {at the
superconducting transition temperature}, $T_c$, as high as
40-150\,Tesla. Wang et al. \cite{wang1} argued that the large
Nernst signal supports a scenario \cite{Kiv}, where the
superconducting order parameter (i.e. the Bogoliubov- Gor'kov
anomalous average $F(\mathbf{r,r^{\prime }})=\langle \psi
_{\downarrow }(\mathbf{{r})\psi _{\uparrow }({r^{\prime }}\rangle
}$)
does not disappear at $T_{c}$ but at much higher (pseudogap) temperature $%
T^{\ast }$. Several other works \cite{others} have also suggested
that the anomalous Nernst effect is a result of the fluctuations
of the superconducting order parameter above $T_{c}$.

However, any phase fluctuation scenario is difficult to reconcile
with the extremely sharp resistive and magnetic transitions at
$T_{c}$ in single crystals of cuprates. The uniform magnetic
susceptibility at $T>T_c$ is paramagnetic, and the resistivity is
perfectly 'normal' showing only a few percent positive or negative
magnetoresistance. Both in-plane \cite{mac,boz,fra,gan} and
out-of-plane \cite {zav} resistive transitions remain sharp in the
magnetic field in high quality samples providing a reliable
determination of the genuine $H_{c2}(T)$. The vortex entropy
estimated from the Nernst signal was found an order of magnitude
smaller than the difference between the entropy of the
superconducting state and the extrapolated entropy of the normal
state obtained by specific heat measurements \cite{cap}. These and
some other observations \cite{lor} do not support any
superconducting order parameter above $T_{c}$.

In this Letter we calculate the Nernst signal for disordered
 conductors with the chemical potential, $\mu$, close to the
mobility edge. No 'Sondheimer cancellation' of the signal exists
in this case. Mott's law \cite{mot} for the variable-range-hopping
conduction of carriers localised below the mobility edge together
with the Boltzmann kinetics for itinerant fermionic carriers or
preformed bosonic pairs above the edge yields the Nernst signal,
which agrees quantitatively with the signal in the superconducting
cuprates  at
 temperatures, $T>T_c(B)$ above the resistive phase transition.

The Nernst voltage is expressed in terms of the kinetic
coefficients $\sigma _{ij}$ and $\alpha _{ij}$ as \cite{ANSE}
\begin{equation}
e_{y}(T,B)\equiv -{\frac{E_{y}}{{\nabla _{x}T}}}={\frac{{\sigma
_{xx}\alpha _{yx}-\sigma _{yx}\alpha _{xx}}}{{\sigma
_{xx}^{2}+\sigma _{xy}^{2}}}},
\end{equation}
where the current density per spin is given by $j_{i}=\sigma
_{ij}E_{j}+\alpha _{ij}\nabla _{j}T$. Carriers in doped
semiconductors and disordered metals occupy states localised by
disorder and itinerant Bloch-like states. Both types of carriers
contribute to the transport properties, if the chemical potential
$\mu$ (or the Fermi level) is close to the energy, where the
lowest itinerant state appears (i.e. to the mobility edge).
Superconducting cuprates are among such poor conductors and their
superconductivity appears as a result of doping, which inevitably
creates disorder. Differently from 3D-conductors, the localised
states cannot be 'screened' off by the itinerant carriers in these
almost two-dimensional conductors even at high density of
carriers. It is well known that in two dimensions a bound state
exists for any attraction, however weak. Indeed, there is strong
experimental evidence for the coexistence of itinerant and
localised carriers in cuprates in a wide range of doping
\cite{tat}.

The standard Boltzmann equation in the relaxation time
approximation yields for itinerant carriers
\begin{equation}
\sigma _{xx}=-e^{2}\sum_{\mathbf{k}}v_{x}^{2}\tilde{\tau}(E_{\mathbf{k}}){%
\frac{\partial f(E_{\mathbf{k}})}{{\partial E_{\mathbf{k}}}}},
\end{equation}

\begin{equation}
\sigma _{yx}=-e^{3}B\sum_{\mathbf{k}}{\frac{v_{x}^{2}}{{m_{y}}}}\tau (E_{%
\mathbf{k}})\tilde{\tau}(E_{\mathbf{k}}){\frac{\partial f(E_{\mathbf{k}})}{{%
\partial E_{\mathbf{k}}}}},
\end{equation}

\begin{equation}
\alpha _{xx}=-e\sum_{\mathbf{k}}{\frac{{E_{\mathbf{k}}-\mu }}{{T}}}v_{x}^{2}%
\tilde{\tau}(E_{\mathbf{k}}){\frac{\partial f(E_{\mathbf{k}})}{{\partial E_{%
\mathbf{k}}}}},
\end{equation}
\begin{equation}
\alpha _{yx}=-e^{2}B\sum_{\mathbf{k}}{\frac{{E_{\mathbf{k}}-\mu }}{{T}}}{%
\frac{v_{x}^{2}}{{m_{y}}}}\tau (E_{\mathbf{k}})\tilde{\tau}(E_{\mathbf{k}}){%
\frac{\partial f(E_{\mathbf{k}})}{{\partial E_{\mathbf{k}}}}},
\end{equation}
where $\mathbf{v}=\mathbf{\nabla }_{\mathbf{k}}E_{\mathbf{k}}$ is the group
velocity, $E_{\mathbf{k}}$ is the band dispersion, $1/m_{i}=\partial ^{2}E_{\mathbf{k}}/\partial k_{i}^{2}$
is the inverse mass tensor, which is assumed to be diagonal, $\hbar =c=1$, $f(E_{
\mathbf{k}})$ is the equilibrium distribution function, and
\begin{equation}
\tilde{\tau}(E_{\mathbf{k}})={\frac{\tau (E_{\mathbf{k}})}{{1+[e\tau (E_{%
\mathbf{k}})B]^{2}/(m_{x}m_{y})}}}.
\end{equation}

Both $\alpha _{xx}$ and $\alpha _{xy}$ vanish at $T=0$ for
degenerate fermions with any $\tau (E_{\mathbf{k}})$, if their
band is parabolic, so that $1/m_{i}$ does not depend on
$\mathbf{k}.$ When $\tau $ does not depend on energy two terms in
the numerator of $e_{y}$, Eq.(1) cancel each other at any
temperature in the parabolic approximation. However, a
generalization of this Sondheimer cancellation for {\it any} band
dispersion is flawed (see, also Ref. \cite{clay2,rom}). The
most striking example is a half-filled band. Modelling this band
by the familiar tight-binding dispersion, $E_{\mathbf{k}}=-2t[\cos
(k_{x})+\cos (k_{y})]$ yields $1/m_{x,y}=\cos (k_{x,y})/m$, where
$m$$=$$1/(2t)$, $t$ is the nearest-neighbour hopping integral,
 and $\mu $$=$$0$ for the half-filling (we take the lattice constant $a=1$).
  Then by parity, $\sigma _{yx}=\alpha _{xx}=0$, but $\alpha _{yx}$ is very large.
  Indeed calculating integrals, Eq.(2) and Eq.(5) we obtain at $k_{B}T\ll t$
\begin{equation}
e_{y}=-{\frac{2t}{{eT\Theta }}}\left( 1-2\Theta \ln ^{-1}{\frac{{1+\Theta }}{%
{|1-\Theta |}}}\right) ,
\end{equation}
where $\Theta =eB\tau /m$. The Nernst signal is negative and
super-linear, $e_{y}\approx -(2t/3eT)(\Theta +4\Theta ^{3}/15)$ at
small $\Theta $$\ll$$ 1$ with  the minimum at $\Theta $$=$$1$. It changes
sign in a strong field, $\Theta $$>$$1$, as shown in Fig.1 inset.  In
this simple example the number of electrons in the lower half of
the band is equal to the number of holes in the upper
half. As a result we arrive  to a substantial $negative$ Nernst voltage, Eq.(7), while both, the Hall effect, $R_H=-B^{-1}\sigma_{yx}/(\sigma_{xx}^2+\sigma_{yx}^2)$ and the thermopower, $S=-\alpha_{xx}/\sigma_{xx}$, equal to zero at {\it any} temperature.
Hence, the Sondheimer cancellation is an exception, rather than a
rule.
\begin{figure}
\begin{center}

\includegraphics[angle=-0,width=0.47\textwidth]{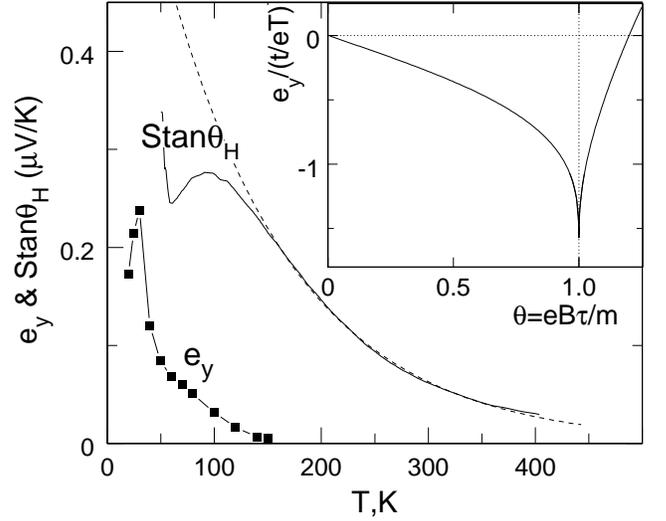}
\vskip -0.5mm \caption{The Nernst signal, $e_y$, and
$S\tan\Theta_H$ in $YBa_2Cu_3O_{6.4}$ at B$=$1Tesla \cite{wang2}.
Inset: $e_y(B)$ in the half-filled band, Eq.(7). }
\end{center}
\end{figure}
However, the thermomagnetic transport in the half-filled band,
Fig.1 inset,  does not describe the experimental results in
cuprates. In particular,  Eq.(7) yields a wrong sign of $e_{y}\approx -60$ $ \mu$V/K and the magnitude, which is at least one order larger than observed with the typical values of $\Theta =10^{-2}$ and $k_{B}T/t=10^{-2}$
\cite{groups,xu,wang1,wang2,cap}. Moreover, in disagreement with the half-filled
band result, the Sondheimer cancellation, $S\tan\Theta_H \gg e_y$, holds in a wide temperature range, as shown in Fig.1 for $YBa_2Cu_3O_{6.4}$. Here $S\tan\Theta_H= \sigma _{yx}\alpha _{xx}/(\sigma _{xx}^{2}+\sigma
_{xy}^{2})$ represents the second term in Eq.(1); S and the Hall angle, $\Theta_H
\approx \tan \Theta_H=BR_H/\rho$, were measured independently. As it is clearly seen from Fig.1,
 $e_y$ and $S\tan \Theta_H$ are of the same order at sufficiently low temperatures,
 also in disagreement with the half-filled band results. Very similar trends of $e_y$ and $S\tan\Theta_H$ were obtained for overdoped LSCO-02 
using independent measurements of $S$, $\rho$, and $R_H$ \cite{LSCO},
in particular, $e_y$ and $S\tan \Theta_H$ are of the same order near $T_c$.

To account for these findings more realistic model is required. When the chemical potential is near the mobility edge,   the effective mass approximation can be applied. In this case, there
is no Nernst signal from itinerant carriers alone, if $\tau $ is a
constant. However, now the localised carriers contribute to the
longitudinal transport, so that $\sigma _{xx}$ and $\alpha _{xx}$
in Eq.(1) should be replaced by $\sigma _{xx}+\sigma _{l}$ and
$\alpha _{xx}+\alpha _{l}$, respectively. Since the Hall mobility
of localised carriers is often much smaller than their drift
mobility \cite{mot}, there is no need to add their contributions
to the transverse kinetic coefficients. Neglecting the orbital
effects($\Theta \ll 1$ \cite {xu,wang1,wang2}) one obtains
\begin{equation}
e_{y}(T,B)={\frac{{{\sigma _{l}\alpha _{yx}-\sigma _{yx}\alpha _{l}}}}{{%
(\sigma _{xx}+\sigma _{l})^{2}}}}.
\end{equation}
When the chemical potential lies near the bottom of the band
($\mu\approx -4t$), $\alpha_{yx}$, Eq.(5), and $\sigma_{yx}$,
Eq.(3),  are positive, but the thermopower of localised electrons
with the energy below $\mu$ is negative, $\alpha _{l}<0$. Hence,
there is no further 'cancellation' in the numerator of Eq.(8) in
this electron-doping regime. When the chemical potential is near
the top of the band ($\mu\approx 4t$), $\alpha_{yx}$ remains
positive, but $\sigma_{yx}$ is negative and $\alpha_l$ is
positive, so that there is no cancellation  in the hole-doping
regime either. In the \emph{superconducting }cuprates the
conductivity of itinerant carriers $\sigma _{xx}$ dominates over
the conductivity $\sigma _{l}$ of localised carriers \cite{tat},
$\sigma _{xx}\gg \sigma _{l}$, which allows us to simplify Eq.(8)
 as
\begin{equation}
\frac{e_y}{\rho}={\frac{k_{B}}{{e}}}r\theta \sigma _{l},
\end{equation}
where $\rho=1/[(2s+1)\sigma_{xx}]$ is the resistivity, $s$ is the
carrier spin, and $r$ is
a constant,
\begin{equation}
\frac{r}{2s+1}=\left({e|\alpha_l|\over{k_B\sigma_l}} +
{{\int_{0}^{\infty} dE E(E-\mu)\partial f(E)/\partial E }
\over{k_BT \int_{0}^{\infty} dE E\partial f(E)/\partial E}}\right
)
\end{equation}
Here  $N(E)$ is the density of states (DOS) near the band edge
($E=0$), and $\mu$ is now taken with respect to the edge.
The ratio $%
e|\alpha _{l}|/k_{B}\sigma _{l}$ is a number of the order of one.
For example, $e|\alpha _{l}|/k_{B}\sigma _{l} \approx 2.4 $, if
$\mu=0$ and the conductivity index $\nu=1$ \cite{end}. Calculating
the integrals in Eq.(10)  yields $r\approx 14.3$ for fermions
($s=1/2$), and $r\approx 2.4$ for bosons ($s=0$).

 The Nernst signal, Eq.(9), is positive, and its maximum value
 $e_y^{max} \approx (k_B/e)r\Theta$   is about
$5$--$10$  $\mu$V/K with $\Theta=10^{-2}$ and $\sigma_l \approx
\sigma_{xx}$, as observed \cite{xu,cap}. Actually, the magnetic
and temperature dependencies of the unusual Nernst effect in the
overdoped LSCO are quantitatively described by Eq.(9), if
$\sigma_l$ obeys the Mott's law,
\begin{equation}
\sigma_l= \sigma_0 \exp \left[-(T_0/T)^{x}\right],
\end{equation}
where $\sigma_0$ is about a constant. The exponent $x$ depends on
the type of localised wavefunctions and variation of  DOS, $N_l$
below the mobility edge \cite{mot,shk,tok}. In two dimensions one
has $x=1/3$ and $T_0 \approx 8 \alpha^2/(k_BN_l)$, where $N_l$ is
at the Fermi level \cite{bry}.

If the magnetic field is strong enough \cite{ref}, the radius of
the 'impurity' wave function $\alpha^{-1}$ is about the magnetic
length, $\alpha $$\approx $$\sqrt{eB}$. If the relaxation time of
itinerant carriers is due to the particle-particle collisions, the
Hall angle depends on temperature as  $\Theta$$ \propto $$T^{-2}$, and the resistivity is linear, $\rho$$ \propto$$ T$ since the density of itinerant carriers is linear in temperature,
both for fermionic \cite{fer} or bosonic (e.g. bipolaronic)
carriers \cite{alebramot}. Hence, the model explains the
temperature dependence of the normal-state Hall angle and
resistivity in cuprates at high temperatures. Finally, using
Eq.(9) and Eq.(11) the Nernst signal is given by
\begin{equation}
{e_y\over{B\rho}}= a(T) \exp\left[-b(B/T)^{1/3}\right],
\end{equation}
where $a(T) \propto T^{-2}$ and $b=2[e/(k_BN_l)]^{1/3}$ is a
constant. Evidently, the phonon drag effect should be
 taken into account at sufficiently
low temperatures in any realistic model. One can account for this
effect by replacing $E_{\bf k}$ in Eq.(4) and Eq.(5) by $E_{\bf
k}+mv^2_s \tau_{ph}/\tau_{e-ph}$ \cite{ANSE}. Here $v_s$ is the
sound velocity,  $\tau_{ph} \propto T^{-4}$ is the phonon
relaxation time due to the phonon-phonon scattering, and
$\tau_{e-ph}$ is the electron (hole) relaxation time caused by
electron-phonon collisions. In two dimensions $\tau_{e-ph} \propto
T^{-1} $ \cite{ale}, so that $a(T)$ in Eq.(12) is enhanced by the
drag effect as $a(T) \propto T^{-6}$. The theoretical field
dependence of $e_y/(B\rho)$, Eq.(12),  is in excellent
quantitative agreement with the experiment, as shown in Fig.2 for
$b=7.32$ (K/Tesla)$^{1/3}$. The corresponding temperature
dependence of $a(T)$ follows closely $T^{-6}$, inset to Fig.2. The
density of impurity states $N_l=8e/(b^3k_B)$ is about $0.4\times
10^{14}$ cm$^{-2}$(eV)$^{-1}$, which corresponds to the number of
impurities  $N_{im}\lesssim 10^{21}$ cm$^{-3}$, as it should be.
\begin{figure}
\begin{center}
\includegraphics[angle=-0,width=0.47\textwidth]{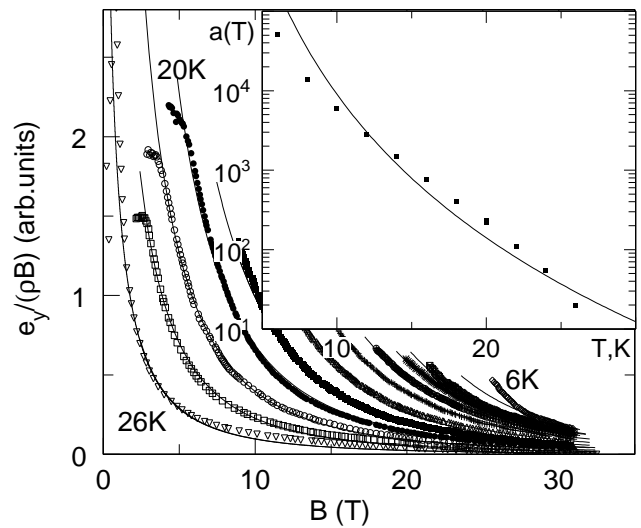}
\vskip -0.5mm \caption{Eq.(12) fits the experimental signal
(symbols) in LSCO-02 
\cite{xu} with $b=7.32$
(K/Tesla)$^{1/3}$. Inset shows $a(T)$ obtained from the fit (dots)
together with $a\propto T^{-6}$ (line). }
\end{center}
\end{figure}

In agreement with the experiment \cite{xu,wang1,wang2}, our  model of
thermal magnetotransport predicts  anomalous Nernst signal in cuprates {\it
only} within the doping interval, where superconductivity is observed. Since  the chemical potential is well below
the mobility edge in the non-superconducting underdoped cuprates
\cite{tat}, and it is  deep inside the Bloch band in heavily doped
samples,
 there is no 'interference' of itinerant and localised-carrier
contributions in these extreme regimes. If carriers are fermions,
then $ S\tan\Theta_H$ should be larger or  of the same order as
$e_y$, because their ratio is proportional to
$\sigma_{xx}/\sigma_{l}\gg 1$ in our model. Although it is the
case in many cuprates (eg., Fig.1, and the text below), a
noticeable suppression of $S\tan\Theta_H$, as compared with $e_y$,
was reported to occur close to $T_c$ in strongly underdoped LSCO
and in a number of Bi2201 crystals \cite{xu}. These observations
could be generally understood if we  take into account that
underdoped cuprates are strongly correlated systems, so that a
substantial part of carriers is (most probably) preformed bosonic
pairs \cite{alebook}. The second term in Eq.(10) vanishes for
(quasi)two dimensional itinerant bosons, because the denominator
diverges logarithmically.  Hence, their contribution to the
thermopower is logarithmically suppressed. It can be almost
cancelled by the opposite sign contribution of the localised
carriers, even if $\sigma_{xx}\gg \sigma_{l}$. When it happens,
the Nernst signal is given by $e_y=\rho \alpha_{xy}$, where
$\alpha_{xy} \propto \tau^2$, Eq.(5). Differently from that of
fermions, the relaxation time of bosons is enhanced critically
near the Bose-Einstein condensation temperature, $T_c(B)$,
$\tau\propto[T-T_c(B)]^{-1/2}$, as in atomic Bose-gases
\cite{bec}. Providing $S\tan\Theta_H \ll e_y$, this critical
enhancement of the relaxation time describes well the temperature
dependence of $e_y$ in Bi2201 and in strongly underdoped LSCO
close to $T_c(B)$. If some segments of a large Fermi-surface
survive in underdoped cuprates, the Bose liquid of preformed pairs
coexists with the fermionic carriers. The degenerate fermions
virtually do not contribute to the thermal transport, but they
dominate  the longitudinal and transverse electric transport.
Hence, the Hall coefficient and resistivity data could not present
a behavior correlated with that of the Nernst signal.

 In conclusion, we calculated the Nernst signal in disordered
conductors with the chemical potential near the mobility edge, and
found   no 'Sondheimer cancellation' of the signal. 'Sondheimer
cancellation' is also absent in the half-filled band, where the
Hall effect and the thermopower are zero, but the Nernst signal is
large and negative. In contrast with the half-filled band, the
model with itinerant and localised fermions and/or charged bosons
 describe quantitatively the anomalous Nernst effect in
high-$T_{c}$ cuprates  as a normal state phenomenon above the
resistive phase transition. Our results strongly support any
microscopic theory of cuprates, which describes the state above
the resistive and magnetic phase transition as perfectly 'normal',
$F(\mathbf{r,r^{\prime }})=0$. Differently from \cite{wang1,wang2}
the present model does not require a radical revision of the
magnetic phase diagram of cuprates \cite{zavkabale}.

 This work was supported by the Leverhulme Trust (grant F/00261/H). We would
like to thank W. Y. Liang and K. K. Lee for helpful discussions.


\begin{thebibliography}{90}
\bibitem{ref} The magnetic field $B\gtrsim 1$ Tesla is strong in
conventional semiconductors with shallow impurity levels
\cite{tok}.
\end{thebibliography}
\end{document}